\documentstyle[preprint, revtex]{aps}
\begin{document}
\draft
\begin{title}
 Event horizons and apparent horizons in spherically symmetric geometries.
 \end{title}
\author{Edward Malec}
\begin{instit}
Physics Department, UCC, Cork, Ireland and

 Institute of Physics, UJ 30-059  Cracow, Reymonta 4, Poland
\end{instit}

\begin{abstract}

 Spherical configurations that are very massive must
be surrounded by apparent  horizons. These in turn,
when placed outside a collapsing body,
have a fixed area and must propagate outward with a
velocity equal to the velocity of radially outgoing
photons. That proves, within the framework of the (1+3)
formalism and without resorting to the Birkhoff theorem,
that apparent horizons coincide with event horizons
in electrovacuum.

\vfill \eject
\narrowtext
\centerline{1. Introduction.}

The Cosmic Censorship Hypothesis (CCH) \cite{11}
 is certainly the most
compelling open problem of classical general relativity.
{}From the dynamical point of view any succesful attempt
to prove a weak version of CCH in a space-time generated
by an isolated self-gravitating object must consist of
the following points:

i) state smooth initial data (if this is done then
  space-time can be unambiguously splitted
into time- and space- directions,initially at least,
 once the local Cauchy problem is solved);

ii) prove that if a singularity emerges then it must
be hidden inside an event horizon, so as not to influence
the asymptotically flat open end;

iii)  prove the  global
Cauchy problem in the asymptotically
flat region outside the event horizon (this implies
that space-time splitting is possible globally in the
region and we are guaranteed the existence of the line
 element  inside it).

{}From this perspective the proof of CCH seems to be
 technically unattainable (at least nowadays) in the
general case of a nonspherical collapse, since the global
Cauchy problem is almost intractable at present
\cite{Christodoulou1}. Even in the spherically symmetric
collapse only partial results are known
\cite{Christodoulou2}.

It is natural to assume the solvability of the global
Cauchy problem, that is to assume the global existence
of a space-time metric, in order to test the
remaining steps of the above programme. In this case
Israel has proven the
confining property of apparent horizons \cite{7}
in spherically symmetric geometries \cite{13}. Israel's
result opens a way to prove CCH for those singularities
 that must be hidden inside apparent horizons. The
formation
of apparent horizons, in turn,
 in spherically symmetric space-times
has been completely solved in the initial value
 formulation \cite{3}, \cite{12}, thus accomplishing
the proof
of steps i) and ii) for a version of CCH in spherically
symmetric geometries.

 \par The intention of this paper is  to provide another
   proof  of the Israel's result  in the
 framework  of a dynamical (1+3) description of a
spherical collapse.  I prove a confining property of
 apparent horizons that are placed in electrovacuum
and complement this with a proof that  a relevant solution
of the global Cauchy problem  exists.

The confining property of apparent horizons in electrovacuum
is not a surprise.
It is well known  thanks to   the Birkhoff theorem
\cite{1}   that once the areal radius  {\bf R}
of a charged collapsing body becomes equal to
$m+\sqrt{m^2-q^2}$, where $m$ is the asymptotic
(Einstein - Freund - Arnowitt - Deser - Misner) mass and
$q$ is the total charge, then the body hides  within an
event horizon that coincides with a sphere of the areal
radius {\bf R} which in turn   is the locus of an
apparent  horizon. This conclusion
appears correct despite the fact  that the Birkhoff
transformation does not exist in the situation of
 interest when a geometry contains apparent horizons
(see a discussion in the Appendix of
\cite{9} in which a generalized
version of the Birkhoff theorem is described).

In the first two sections I will assume the existence
of a global Cauchy solution, that is the existence
of an asymptically flat  space-time
with  the spherically symmetric metric line element
\begin{equation}
ds^2=-\alpha^2(r, t)dt^2 +a(r, t)dr^2
+b(r, t)r^2d\Omega^2.\label{1}
\end{equation}

$r$ is a coordinate radius and $br^2 d\Omega^2 $
is a standard 2-sphere metric element.
 We assume   the maximal gauge condition  in which
  components of the extrinsic curvature $K_{ij}$ of
the hypersurface $\Sigma_t $  (defined as a set of points
having a fixed coordinate time $t=const$)
satisfy the following equations

\begin{equation}
K^r_r={{\partial_t a}\over {2a\mid \alpha \mid }}=
-2K^{\phi}_{\phi }= -2K^{\theta }_{\theta }=
 -{{\partial_tb}\over b\mid \alpha \mid }. \label{2}
\end{equation}
We use the standard convention \cite{1} in which
quantities supplied with Greek type indices refer to
the four-geometry while quantities with Latin
labels refer to the geometry of the hypersurface
$\Sigma_t$. The Einstein summation
convention is applied in some places, with the exception
concerning the label $r$, whose repetition
is assumed never to mean summation. Below
$\nabla_i $ denotes the covariant derivative on $\Sigma_t $.
The Einstein equations read

\begin{equation}
-{2\over abr^2}\partial_r(r^2\partial_rb)
+{(\partial_rb)^2\over 2ab^2}+{\partial_rb\partial_ra
\over a^2b}+{2\partial_ra\over a^2r}-{2\partial_rb\over
abr} +{2\over r^2}({1\over b}-{1\over a})=
K_{ij}K^{ij} +16\pi \rho .
\label{1a}
\end{equation}

\begin{equation}
\nabla_iK^{ir}= -8\pi j^r . \label{1aa}
\end{equation}

\begin{equation}
\nabla_i\partial^i\alpha = K_{ij}K^{ij}\alpha +
4\pi [-T_0^0 +T_i^i]\alpha
\label{1aaa}
\end{equation}

\begin{equation}
\partial_t K^r_r = -\alpha  R^{(3)r}_r -8\pi \alpha
[-T_r^r +{T_{\mu }^{\mu }\over 2}] +\nabla_r\partial^r
\alpha
\label{1aaaa}
\end{equation}

Above $T_{\mu }^{\nu }$ is the energy momentum tensor
of matter generating the gravitational field, $\rho =
-T^0_0$ and $j_i = -T^0_i \alpha $. $R^{(3)r}_r$ denotes a radial
component of the 3-dimensional Ricci tensor which might
be expressed (using the hamiltonian constraint) as follows

\begin{equation}
R^{(3)r}_r=-8\pi T_0^0 +
{K_{ij}K^{ij}\over 2} +{(\partial_rb)^2\over 4ab^2}
+{\partial_rb\over abr} -{1\over br^2}+{1\over ar^2}
.\label{1aaaaa}
\end{equation}

The equations (\ref{2},\ref{1aaaa}) are dynamical ones
while (\ref{1a}, \ref{1aa}) are the hamiltonian and
 momentum
constraints \cite{1},  respectively. As initial data one
may take
for instance $a(r,0)=b(r,0)$ at a time $t=0$; given the
 matter
distribution one obtains from (\ref{1a}, \ref{1aa}) the
three-geometry of $\Sigma $ and the extrinsic
curvature $K_r^r$ and from (\ref{1aaa})
the lapse $\alpha $. Equations (\ref{2}, \ref{1aaaa})
    determine the rate of change of the
three-geometry   and of $K_r^r$.
Of the two functions $a, b$ only one is
independent. In the maximal slicing $\partial_t(b^2a)=0$;
taking into account the above initial condition one gets

\begin{equation}
b^2(r,t)= {b^3(r,0)\over a(r,t)}.
\label{1ab}
\end{equation}

The order of the rest of this paper is following.
In Section 2 it is proved that the apparent horizon
is null-like in electrovacuum, has a constant area
and coincides with
 the event horizon. Section 3 contains
a proof of a version of the global Cauchy problem.
In the last Section I comment on the
significance of the results and their (possible)
generalization.
\vskip 2cm
\centerline{2. Main calculations.}
\vskip 1cm

Let the boundary of a collapsing body be a sphere of
a coordinate radius $r_0$. Let us define  $c=
{{-q^2+m^2}\over 4}$.
First, let us notice that there exists
a  solution (Appendix 1) that is manifestly
static outside a collapsing body \cite{2}
\begin{equation}
a=b=(1+{m\over r}+{c\over { r^2}})^2,\label{1b}
\end{equation}

\begin{equation}
\mid \alpha  \mid =
 \bigg| {{r-\sqrt{c}}\over {r+\sqrt{c}}}
\bigg|^{{m\over {2\sqrt{c}}}}~\bigg| r^2-c\bigg|^{{{q^2}
\over {2(r^2+mr+c)}}}~e^{{{q^2}\over 2}\int_r^{\infty }
{{\ln (\mid s^2-c^2\mid )(2s+m)}\over {(s^2+ms+c)^2}}ds}
\label{1bb}
.\end{equation}

$\alpha $ vanishes at $r=\sqrt{c}$ quicker than
$r-\sqrt{c}$, since ${m\over {2\sqrt{c}}}\ge 1$. From
this one readily infers that at a surface  $S$ placed
at a coordinate radius $r=\sqrt{c}$ (that is, at an
areal radius  ${\bf R}=m+\sqrt{m^2-q^2}$) exists an
event horizon; no signal can traverse through  $S$ in
 a finite coordinate  time $t$.  In the case of  vanishing total
 charge $q$ the  corresponding line element  coincides
 with the Schwarzschild  line element  in isotropic
coordinates  \cite{1}.  The solution (\ref{1b}, \ref{1bb})
will not be considered in the rest of this paper.

\par Let us assume that a part of a  spherically
symmetric spacetime generated by a collapsing body can
be foliated by maximal slices $\Sigma_t$, $0\le t\le
 t_0$ that are asymptotically flat. Assume also that
there exists a smooth continuation of the above band of
slices, hypersurfaces $\Sigma_t^{out}$, that are maximal
outside a region of compact support and that cover
a region with  the outermost apparent horizon (if it
exists).   The coordinate time $t$ is  a parameter that
labels maximal slices but it coincides with a proper time
 of an external
observer that is localized very far from a collapsing
body.

 It was proven elsewhere \cite{3} that when the amount
 of matter minus a total radial momentum exceeds a
multiple ( 1 or 7/6)  of the proper radius, then
apparent  horizons must form. Let us  assume that
there exists an apparent horizon outside a  (neutral or
charged) collapsing body of  a compact support. (We
do not exclude electrovacuum, i. e.,
there   might exist  longe-ranged potentials
 outside a body, with an  electrostatic Coulomb-like
energy density.)

Under these conditions, one proves that the
Penrose \cite{GG}, \cite{5} inequality (which
 actually becomes an equality) holds true, that at
the surface of an apparent horizon

\begin{equation}
m = \sqrt{S\over {16\pi }} +
q^2\sqrt{{\pi }\over S}.\label{3}
\end{equation}

\par
It will be convenient to prove (\ref{3})
in an isotropic system of coordinates in which
$a=b=\phi^4$. It is easy to prove that this form of a
metric can be achieved just by performing
a suitable change of a radial coordinate on a fixed
Cauchy slice. Morever, the final result - equation
(\ref{3}) - is  already expressed in a coordinate
independent way.

 The proof goes as follows.
In electro-vacuum the hamiltonian  constraint reads
\cite{6}
\begin{equation}
\hat \Delta \phi = -{1\over 4}\hat E_i\hat E^i \phi^{-3}
- {{ K_{ij} K^{ij}\phi^5} \over 8}.\label{4}
\end{equation}
Here the hatted quantities refer to the flat background
metric and $\hat E^r = {q\over {r^2}}, E^{\theta }=
E^{\phi }=0.$ From the momentum constraints one gets
\cite{3}

\begin{equation}
K_{ij}= (n_in_j-{{g_{ij}}\over 3}){C\over
{\phi^6R^3}}, \label{5}
\end{equation}

where $n_{j}$ is a normal vector in the physical
non-hatted metric. C depends on time but
it  is   constant on a part of a fixed Cauchy slice
 that is exterior to the collapsing body. \par
Equation (\ref{4}) has a conserved ($r-$independent,
 $\partial_r E=0$) quantity
\begin{equation}
E={r\over 8}(2r\partial_r\phi +\phi )^2-
{{r\phi^2}\over 8} - {{q^2}\over {8r\phi^2}}-
{{C^2}\over {72 r^3 \phi^6}}. \label{6}
\end{equation}
Assuming asymptotic flatness one finds that $E=-{m\over
4}$, where $m$ is the asymptotic
 mass.
Notice also that $r\phi^2$ is equal to an areal
radius, $r\phi^2=\sqrt{{S\over {4\pi }}}$.
 After some rearrangements one might
write (\ref{6})  as follows
\begin{equation}
m - [\sqrt{S\over {16\pi }} +
q^2\sqrt{{\pi }\over S}] =-{r\over 2}[2r\partial_r \phi
+\phi  -{C\over {3r^2\phi^3}}][2r\partial_r \phi
+\phi  +{C\over {3r^2\phi^3}}]=
-{{{\bf R}^3}\over 8}\theta (S)\theta '(S).\label{7}
\end{equation}

Here $\theta (S)$ is the divergence
of outgoing light rays,
$\theta (S) = {1\over {\alpha }}
{d\over {dt}}\ln S$ (the
derivative in the direction of
outgoing photons) and
 $\theta '(S) = -{1\over {\alpha }}{d\over {dt}}\ln S$
(the
derivative in the direction of intgoing photons)$^7$
is the convergence of ingoing light rays.  If $S$ is
an apparent horizon, then  $\theta (S)$ vanishes
which proves (\ref{3}). \par  Equation (\ref{3})
actually holds on all  maximal slices or on  their
 time developments that are maximal  in an  asymptotically
flat  region  (up to the apparent
horizons) as far as the apparent horizons remains
 outside the collapsing body. That means that the areal
radius  \begin{equation}
{\bf R}=r\sqrt{b}\label{8}
\end{equation}
(we are coming back to the original metric notation
(\ref{1})) of the apparent horizon must be conserved in
time, since the mass $m$ is  conserved in time in
asymptotically flat systems. That is, the full time
derivative of {\bf R} must vanish, which leads to the
following equality
\begin{equation}
V\sqrt{b}{{2b+r\partial_r b}\over {2b}}+
r{{\partial_tb}\over {2\sqrt{b}}}=0.\label{9}
\end{equation}

Here $V=dr/dt$ is the coordinate velocity expansion of
the apparent horizon.
Now, the condition for the apparent horizon reads
\begin{equation}
2b+r\partial_r b={{brK_{rr}}\over {\sqrt{a}}};\label{10}
\end{equation}

inserting this into (\ref{9}) and using (\ref{2}) we
obtain

\begin{equation}
{{\sqrt{b} K_{rr}r}\over {1\sqrt{a}}}
[V- {{\mid \alpha \mid }\over\sqrt{a}}]=0.\label{11}
\end{equation}
We conclude that
the apparent horizon expands with a radial velocity
\begin{equation}
V={{\mid \alpha \mid }\over {\sqrt{a}}}.\label{12}
\end{equation}
But, from  equation (1) we know that
${{\mid \alpha \mid}\over {\sqrt{a}}}$ is equal to the
velocity of radially outgoing photons. Therefore, no
material object can escape from within the
apparent horizon; it just coincides with  the event
horizon. We can say more. Actually, the equation
(\ref{7}) might be interpreted in the following
way: if an areal radius of a sphere $S$ satisfies
(\ref{3}), then $S$ is an apparent horizon. By
 continuity, all Cauchy slices must contain surfaces
satisfying (\ref{3}), and the surfaces
 must remain in vacuum,
since they move with the velocity of light.
That means that, as long as maximal slices exist (or
 a time development of an initially maximal slice
that is partially maximal later on, that is it remains
maximal  in  an  open end containing the
apparent horizon) and under
conditions stated previously, the apparent horizon must
exist forever and it coincides with an event horizon.
In this way we have proven in the framework of (1+3)
formalism, without resorting to the Birkhoff theorem,
 that there exists an event horizon that intersects each
Cauchy slice along a sphere of the areal radius
${\bf R}= m+\sqrt{m^2-q^2}$.

Let us remark that one can prove in a similar way,
that when a locus of points with $\theta '=0$ (i. e.,
the past  apparent horizon) is placed in electro-vacuum,
 then  it moves to the centre with the velocity of light.
This set of points is impenetrable from outside and its
area is constant, hence it constitutes a boundary of a
white hole \cite{8}. The solution (\ref{1b}, \ref{1bb}) is
simultaneously a white hole and a black hole since
$\theta =\theta '=0$ at $r=\sqrt{c}$. The velocity of
  the  boundary $r=\sqrt{c}$ is  equal to zero.

There exists yet another possibility to prove that if
the evolution of a collapsing  matter is smooth outside
 a region of compact support  (the  latter can contain
singularities),   then if an apparent horizon exists at
a time $t$ then it must exist forever. This is obvious,
if we notice that the  development of the divergence
$\theta $ in the direction of outgoing photons is given
by the Raychaudhury equation \cite{9}, which    in
electrovacuum reads in (1+3) splitting \cite{10}
\begin{equation}
{d\over {dt}}\theta = {{\theta }\over
\sqrt{a}}(\partial_r\alpha +\alpha \sqrt{a} K_r^r)
-{{\alpha \theta^2}\over 2}
\label{15a}.
\end{equation}
{}From  (\ref{15a}) one can deduce that a surface with
vanishing divergence $\theta $ propagates to the future.

Equation (\ref{3}) provides a well known necessary and
(simultaneously)  sufficient condition for the
formation of event horizons formulated in terms of
asymptotic quantities $m, q$ and an area $S$. One
can obtain also  criteria in terms of quasilocal
quantities, by combining  results of this paper with
 some  of theorems of \cite{3}. One can, for instance,
formulate the following statement:
\par {\bf Theorem 1.} If on an initial spherically
symmetric Cauchy hypersurface with momentarily static
initial data

\begin{equation}
M>L,\label{13}
\end{equation}
where $M$ is a total mass \cite{3} and $L$ is a proper
radius of a collapsing body, then the whole space-time
history of the body must contain an event horizon that
surrounds the body.  \par Thus, if the energy content
inside a ball of a fixed radius becomes large, then it
hides under an event horizon. That proves a version of
the Cosmic Censorship Hypothesis \cite{11} (CCH) in
which $singularities$ are supplied with a qualifier
$massive$; {\bf massive singularities are hidden under
an event horizon}- this is a version of CCH that looks
plausible.

The formulation in terms of quasi-local quantities is
of interest, since it can be pursued further to cover
cases in which  the standard approach fails.  In
spherically symmetric geometries (asymptotically flat
and in some cosmological models) event horizons must
exist if   apparent horizons  are present   \cite{10}.
The quasi-local conditions that imply the formation of
apparent horizons in spherically symmetric geometries
are already known \cite{3}, \cite{12}.   Using them, one
can  obtain a number of conditions for the formation of
event horizons inside collapsing matter (in asymptotically
flat geometries)  and in cosmological models.

\vskip 2cm

\centerline{3. The Cauchy problem.}

\vskip 1cm

In the above considerations I have assumed the existence
of  a global maximal  Cauchy surface  which possesses
a maximal extension at least in the part of a space-time
that is exterior to the apparent horizon and
which  includes the latter. Let me point out that in
standard  proofs  of the Birkhoff theorem one usually
assumes the existence of that part of space-time that is
exterior to the collapsing body; that is  merely
equivalent  with my conditions. Nevertheless there exists
a possibility to get rid of the assumption. Below I  sketch
a line of reasoning that should lead to a proof of a
 version of the global Cauchy problem.

To pursue further
we will need the spherically symmetric Einstein equations
in electrovacuum. In  electrovacuum some of
 the matter related terms (i. e., $j_r, T^{\mu }_{\mu }$)
of equations (\ref{1aa}, \ref{1aaaa} )  vanish.
 Notice that   $K_{ij}K^{ij}$
 can be written   (due to spherical symmetry)  as
${3\over 2}(K_r^r)^2$.
 Below appears  the mean curvature $p$
of a sphere as embedded
in a hypersurface $\Sigma_t $

\begin{equation}
p={(r\partial_rb +2b)\over \sqrt{a}br }.
\label{a0}
\end{equation}

The energy density $\rho $  contains only  a contribution
  from the electrostatic  field, $\rho = {q^2\over
8\pi r^4b^2}={q^2\over 8\pi R^4}$.

The   Einstein equations read

\begin{equation}
-{2\over abr^2}\partial_r(r^2\partial_rb)
+{(\partial_rb)^2\over 2ab^2}+{\partial_rb\partial_ra
\over a^2b}+{2\partial_ra\over a^2r}-{2\partial_rb\over
abr} +{2\over r^2}({1\over b}-{1\over a})=
{3\over 2}(K_r^r)^2 + 16\pi \rho ,
\label{a1}
\end{equation}

\begin{equation}
\partial_rK^r_r= -{3\sqrt{a}\over 2} p K^r_r , \label{a2}
\end{equation}

\begin{equation}
{1\over {a^{1/2}br^2}}\partial_r(a^{-1/2}br^2\partial_r
\alpha )=
[{3\over 2}(K_r^r)^2 +8\pi \rho ]\alpha  ,
\label{a3}
\end{equation}

\begin{equation}
\partial_ta= 2\alpha a K_r^r ,
\label{a4}
\end{equation}

\begin{equation}
\partial_tK^r_r = {3\alpha  \over 4}(K_r^r)^2  -
{\alpha p^2\over 4}-{p\partial_r \alpha
\over \sqrt{a}}+{\alpha \over br^2}-8\pi \alpha \rho .
\label{a5}
\end{equation}

\vskip 0.5cm
\centerline{3a. The initial data.}
\vskip 0.5cm
Assume that $\Sigma_0$ is a global Cauchy hypersurface.
As pointed out  at the end of section 1,  initial
data are determined by the initial distribution of matter
and momentum; without  loss of generality we can
assume that initially $a(r)=b(r)$. If $\Sigma_0$ contains
an apparent horizon, then a singularity shall develop
in the future, therefore the best we can hope to prove
 is the existence of a solution outside the apparent
horizon. We wish to consider the Cauchy problem outside
the apparent horizon; this seems to be reasonable, since
 the apparent horizon moves outward with the speed of
light and nothing that happens inside it can casually
influence its exterior.

Let $\Sigma_t^{out}$ be a Cauchy maximal hypersurface
that evolves from $\Sigma_0$ in the region outside the
cylinder enclosed by the apparent horizon.
 $\Sigma_0^{out}$ concides obviously with a corresponding
part of $\Sigma_0$, hence  the Cauchy data
   $K_r^r$ and $a$ are fixed.  (The function $b$ might be
 determined from (\ref{1ab}) and is not an independent
dynamical quantity.)
 The hamiltonian constraint constitutes an elliptic
equation. From asymptotic flatness we have to set
$a(\infty )=b(\infty )=1$, but this condition is
not sufficient to ensure the uniqueness of solutions
of equation (\ref{a1}) on $\Sigma_t^{out}$.
 However, the
asymptotic mass $m$ must be constant on all slices and it
is determined by the geometry of $\Sigma_0$.
 Therefore we must demand that on all Cauchy slices
 $\Sigma_t^{out}$ the following asymptotic
condition is met

\begin{equation}
  \lim_{r\rightarrow \infty }
r^2\partial_ra=\lim_{r\rightarrow \infty }
r^2\partial_rb=-2m
\label{a6}
\end{equation}.

In what follows we assume that    the convergence
 $\theta ' =K_r^r+p$  of ingoing light rays is
strictly positive on $\Sigma_0$; this implies that
 $\theta ' =K_r^r+p$ is strictly positive on all future
slices $\Sigma_t^{out}$ (Appendix 2).
  If there is an apparent horizon, then
$\theta (r)=K_r^r-p$ vanishes at a centered sphere in
 $\Sigma_t^{out}$; the preceding  assumption
$\theta ' >0$ implies that
$K_r^r > 0 $ at the apparent horizon and out of it, since
$K_r^r$ does not change sign in electrovacuum (see formula
(\ref{5})). Now we can conclude that at the outermost
apparent horizon and out of it we must have $p > 0$, i.e.,
there is no minimal surface in  $\Sigma_t^{out}$.
Let $r=r_{t}$ be a position of the outermost apparent
 horizon.
We may invoke the calculation  performed in the
previous section, which led to the  equation (\ref{7}).
(The calculation bases on the assumption that $a=b=\phi^4$
but it is only a technical condition and there is no
any loss
of generality.)  At the apparent horizon (\ref{7}) yields
\begin{equation}
m - [\sqrt{S\over {16\pi }} +
q^2\sqrt{{\pi }\over S}] = 0.\label{7a}
\end{equation}
 In electrovacuum there are two solutions of the
equation (\ref{7a}),

\begin{equation}
r_t\phi^2(r_t)|_2^1= m^+_-\sqrt{m^2-q^2};
\label{7b}
\end{equation}

taking into account the above conditions  we have to
choose

\begin{equation}
\phi (r_t)= \sqrt{{1\over r_t}(m+\sqrt{m^2-q^2})}
\label{7c}
\end{equation}

since otherwise there could exist a minimal surface at
some $r >r_t$. The hamiltonian constraint  is now supplied
with the standard Dirichlet boundary conditions
$\phi (r_t), \phi (\infty )=1$ on
$\Sigma_t^{out}$ and it is easy to prove that  there is
 a  unique solution.

Thus, under the above conditions fixing the asymptotic
mass $m$ uniquely determines the conformal factor $\phi $
(and, consequently, $a, b$, if a relation between the two
functions is determined on an initial slice)
 at the surface of the apparent horizon. But $m$ is
determined by the initial geometry $\Sigma_0$, hence
 we have no any freedom left in specifying
the solutions of the hamiltonian constraint in the
exterior region.

 The corresponding boundary problem for
the lapse equation (\ref{a3}) contains, however, an
arbitrariness. The condition

\begin{equation}
\alpha (\infty )=1
\label{a7a}
\end{equation}

does not specify uniquely a solution of (\ref{a3});
we still can impose

 \begin{equation}
\partial_r\alpha  = f(t)
\label{a7}
\end{equation}

at the surface of an outermost  apparent horizon.
I assume that the function $f(t)$ is smooth and
strictly positive.  We have defined
the exterior Cauchy problem (i. e., with data
on $\Sigma_t^{out}$) as   a restriction
of the global Cauchy  problem with data on $\Sigma_t$.
   There is an obvious loss of information
during such a restriction, since we do not control
the collapse of matter fields that are enclosed
inside the apparent horizon.
The arbitrariness in choosing
$f(t)$  corresponds to our unawareness about the
full state of the collapsing system.

In summary, the initial value problem of the
electrovacuum Einstein equations outside the
apparent horizon  can be  determined by prescribing
the asymptotic mass $m$, a relation $a(r,t=0)=b(r,t=0)$,
 an initial
datum  $ K_r^r$ at a time $t=0$ and a condition
(\ref{a7}) for the lapse function.

\vskip 0.5cm
\centerline{3b. The exterior Cauchy solution
exists globally.}
\vskip 0.5cm
{\bf Theorem 2.} Let
$H_s(\Sigma_{t}^{out})$ be a Sobolev space of
functions defined
on $\Sigma_t^{out}$. Let $\Sigma_{t=0}$ be a
global maximal
Cauchy
surface with initial data $a(t=0)=b(t=0)$ such that
 $\partial_ra(r,t=0)
\epsilon H_{2}(\Sigma_{t=0})$
, $K_r^r(t=0)\epsilon H_2(\Sigma_{t=0})$
generated by a given initial distribution of matter.
Assume that  the convergence $\theta '(S)$
of the ingoing light rays is strictly positive for any
centered sphere $S$ on $\Sigma_0^{out} $ and
that there exists
an apparent horizon that is placed in vacuum or in
electrovacuum. Let the lapse function
satisfy the boundary conditions (\ref{a7a}, \ref{a7}) with
 $f(t) \ge 0$ at the outermost apparent horizon.
 Then there exists a unique  solution of the
global Cauchy problem in the region exterior to the
apparent horizon, including the apparent horizon itself.

\vskip 0.5cm
{\bf Sketch of the proof.}

There exist theorems \cite{13a} from which one infers the
 existence of a  solution ($K_r^r\epsilon H_s(\Sigma_t),
\partial_r a \epsilon H_s(\Sigma_t)$
or $K_r^r, \partial_r a \epsilon H_s^{loc}(\Sigma_t)$ )
for $s\ge 2$ and sufficiently short intervals of time.
The lower bound on the index $s$ is due to the
 Scha\"uder ring
property \cite{13ab} which is fulfilled  in three spatial
dimensions if $s\ge 2$.
 Assuming the local existence, I will estimate  (Step 1)
the pointwise
growth of $a(t)$, $K_r^r(t), \partial_rK^r_r,
p, \partial_rp, \partial_r\alpha $. Using those estimates
and the method of energy estimates
\cite{13b}   one can  show (Step 2) that
Sobolev norms in question ($K_r^r\epsilon
H_s(\Sigma_t^{out}), \partial_ra
\epsilon H_s(\Sigma_t^{out})$)
do not blow up in a finite
time $t$, thus accomplishing the final goal.
\vskip 0.5cm
{\bf Step 1. The $L_{\infty }$ estimates.}
\vskip 0.5cm
In electrovacuum the right hand side of  (\ref{a3})
is nonnegative. Invoking the maximum principle and using
the boundary condition  that the lapse satisfies at the
apparent horizon we conclude that

\begin{equation}
\partial_r\alpha \ge  f(t) \ge 0
\label{3.1}
\end{equation}

everywhere in $\Sigma_t^{out}$.
As pointed out in the subsection (3a), if $\theta ' >0$
then

\begin{equation}
 p > K_r^r>0
\label{3.2}
\end{equation}

everywhere in $\Sigma_t^{out}$ (but at spatial infinity
both functions vanish, $K_r^r =O(1/r^3)$ and $p=0(1/r)$)).
 Notice also, that

\begin{equation}
a(r,t)=a(r,0)+2\int_0^t\alpha K_r^rds\ge a(r,0)=
b(r,0)\ge b(r,0)- 2\int_0^t\alpha K_r^rds=b(r,t)
\label{3.3}
\end{equation}.

Using the inequalities (\ref{3.1}, \ref{3.2})
one gets from  (\ref{a2}) and (\ref{a5})

\begin{equation}
(\partial_t+{\alpha \over \sqrt{a}}\partial_r)
K^r_r \le {1  \over br^2}
  \le
 {1 \over R_{AH}^2},
\label{3.4}
\end{equation}

where I used $\alpha \le 1$ and the obvious fact
 (since $p>0$ on $\Sigma_t^{out}$) that the areal
 radius $R=\sqrt{b}r$ is bounded from below by the
areal  radius of the apparent horizon $R_{AH}=
 m+\sqrt{m^2-q^2}$. Let $S^*$  be an intersection of
$\Sigma^{out}_0$ with a light cone passing through a
sphere of a coordinate radius $r$ in $\Sigma^{out}_t$
and let $r^*$ be a coordinate radius of $S^*$.

 $K_r^r(t,r)$   is  bounded from above,

\begin{equation}
 K_r^r(r,t) \le K_r^r(r^*, 0) + {t\over R^2_{AH}}.
\label{3.5}
\end{equation}

 The equation (\ref{a2}) can be solved on each slice
$\Sigma_t^{out}$ to give

\begin{equation}
K_r^r(r,t)= {C(t)\over  R^3}.
\label{3.6}
\end{equation}

  At the apparent
horizon $p=K_r^r$ and   $\theta '=p-K_r^r$
is positive during the collapse (Appendix 2)
if it is positive on $\Sigma_0^{out}$; this implies that
$K_r^r$ is positive at the horizon and from (\ref{3.6})
also $C(t)$ is positive.

  The extrinsic curvature
$K_r^r$  decreases on a fixed slice $\Sigma_t^{out}$ and
achieves its largest value at the apparent horizon.
 The chain of inequalities

\begin{equation}
{C(t)\over  R^3}\le {C(t)\over  R^3_{AH}}
\le {C(0)+tR_{AH}\over R^3_{AH}}
\label{3.7}
\end{equation}

leads to the final estimations

\begin{equation}
C(t) \le C(0)+tR_{AH}, ~~~K_r^r(r,t)\le {C(0)+tR_{AH}
\over R^3}.
\label{3.8}
\end{equation}

(\ref{2}) and (\ref{3.8}) imply an    estimation

\begin{equation}
\partial_t a(r,t)= 2a(r,t)\alpha (r,t)K_r^r(r,t)\le
 a(r,t) 2{C(0)+tR_{AH}\over R^3_{AH}}
\label{3.9}
\end{equation}

 which in turn gives the following
estimation of the metric coefficient $a(r,t)$:

\begin{equation}
a(r,t)\le a(r,0) e^{  {2tC(0)+t^2R_{AH}
\over R^3_{AH}}}.
\label{3.10}
\end{equation}

A strightforward calculation gives

\begin{equation}
\partial_t p(r,t)= -{ \partial_r\alpha (r,t)\over
\sqrt{a}}K_r^r(r,t) +{\alpha K_r^r\over 2}p\le
{\alpha K_r^r\over 2}p\le   {C(0)+tR_{AH}
\over 2R^3_{AH}}p;
\label{3.11}
\end{equation}

above I employed the inequalities $\partial_r\alpha
\ge 0$,
$\alpha \le 1$ and (\ref{3.8}).

One can show also that

\begin{equation}
\partial_r p(r,t)=
-\sqrt{a}(8\pi \rho + {3\over 4}(K_r^r(r,t) )^2
 +{3\over 4}p^2 -{1\over br^2}).
\label{3.11a}
\end{equation}

(\ref{3.11}),   (\ref{3.11a}) and the previous estimates
  yield

\begin{equation}
(\partial_t+{\alpha \over \sqrt{a}}\partial_r )p=
8\pi \alpha ({j_r\over \sqrt{a}}-\rho )-{\partial_r\alpha \over \sqrt{a}}
K_r^r
-{\alpha \over 4}[ \theta^2+2(K_r^r)^2 +2p^2]
+{\alpha \over R^2}\le
{1\over R^2},
\label{3.11c}
\end{equation}

which gives finally

\begin{equation}
p(r,t)\le \sup [p(r, t=0)] +{ t \over R^2_{AH}}.
\label{3.12}
\end{equation}

A combination of (\ref{3.8}, \ref{3.9}) and  (\ref{3.12})
together with the momentum constraint (\ref{a2})
allows one
to obtain a bound for $\partial_rK_r^r(r,t)$:

\begin{equation}
\mid {\partial_r K_r^r(r,t)\over \sqrt{a(r,t)}} \mid \le
 {3(C(0)+tR_{AH})\over R^3}(\sup [p(r,0)]+{t\over
 2R^2_{AH}}).
\label{3.11b}
\end{equation}

The energy density is positive and  bounded from  above
by ${q^2 \over 8\pi R^4_{AH}}$; using the
preceding information one easily infers that
$\mid \partial_rp\mid $ must be uniformly
bounded on all slices
$\Sigma_t^{out}$ by $C_1 t^2+C_2t +C_3$ where $C_1,
C_2, C_3
$ are  constants depending only
on the initial data.

 In order to get estimations  of $\partial_r\alpha
 (r,t), \partial_r^2 \alpha (r,t)$ one should
analyze the lapse equation (\ref{a3}) using the above
information about the evolution of the extrinsic
curvature.
A lengthy and not particulary illuminating calculation
gives  a bound

\begin{equation}
({\partial_r\alpha \over \sqrt{a}})(r,t)\le
f(t) +
{4\pi q^2 \over RC(t)}+{3\pi C(t)\over R^2R_{AH}}).
\label{3.13}
\end{equation}

The right hand side of (\ref{3.13}) is finite since
$C(t)$ is strictly greater than 0 in any finite time $t$
(Appendix 2) and satisfies (\ref{3.8}).

In summary, the following estimates hold true

\begin{equation}
||X ||_{L_{\infty }(\Sigma_t^{out})}\le
 F(t)||X ||_{L_{\infty }(\Sigma_0^{out})},
\label{3.14}
\end{equation}

where $X$ is any of the  functions ($a,
K_r^r, \partial_rK_r^r, p,   $) and $F(t)$ is a   positive
function that remains bounded for arbitrarily large but
 finite values of its argument. The lapse $\alpha $
does not exceed 1, $\partial_r\alpha $ must satisfy
(\ref{3.13}) and $\mid \partial_rp\mid $ is bounded
by a quadratic function of $||X ||_{L_{\infty }(\Sigma_t^{out})}$.

\vskip 0.5cm

{\bf Step 2. The integral estimates.}

\vskip 0.5cm
In order to prove the existence of a global solution
one has to show that the Sobolev norm
$||K_r^r||_{H_2(\Sigma_t^{out})}+
||\partial_ra||_{H_2(\Sigma_t^{out})}$ remains bounded
for any finite time $t$. This could be done explicitly,
by pursuing the above calculation in order to get
pointwise estimates for all quantities in question
and then
proving the required integrability. I will choose
a way that is probably less economic in this
particular case but offers a chance for generalization.

 Let ${\bf X}$ be a vector
having 6 components ${\bf X}_i=\partial_r^i a,
{\bf X}_{i+3}=\partial^{i-1}_rK_r^r$, $i=1, 2, 3$.
Define
 \begin{equation}
 H(t) = ( ||K_r^r||^2_{H_2(\Sigma_t^{out})}+
||\partial_ra||^2_{H_2(\Sigma_t^{out})}).
\label{3.15aa}
\end{equation}
A simple but laborious calculation shows that

\begin{equation}
 {d\over dt} H=
\sum_{i,j=1}^6\int_{\Sigma_t^{out}}dV {\bf X}_i{\bf X}_j
f_{ij}+\sum_{i=1}^6f_iX_i - 4\pi \alpha R^2
\sum_{i=1}^6X_i^2\mid_{AH},
\label{3.15}
\end{equation}

where $f_i, f_{ij}$ are certain  polynomials of finite
order  that depend
 on $\sqrt{a}, b, K_r^r, \partial_rK_r^r, p, \partial_rp,
\partial_r\alpha , \alpha ,\rho $. The functions $f_i$
are square integrable.
The crucial point is that the right hand side of
(\ref{3.15}) is bilinear in ${\bf X}_i$.
Direct differentiation of $||K_r^r||^2_{H_2
(\Sigma_t^{out})}+
||\partial_ra||^2_{H_2(\Sigma_t^{out})}$ with
respect $t$ and the use of evolution equations
(\ref{a4}, \ref{a5}) gives also some trilinear terms but
   manipulating  with
equations (\ref{a1}- \ref{a3}) and   (\ref{1ab})
finally yield the above equation.

The right hand side of (\ref{3.15}) is bounded from above
by

\begin{equation}
\sup \lbrace ||f_{ij}||_{L_{\infty }(\Sigma_t^{out})}
 \rbrace \sum_{i=1}^6 \int_{\Sigma_t^{out}}dV{\bf X}_i
{\bf X}_i+
\sum_{i=1}^6 [\int_{\Sigma_t^{out}}dVf_i^2]^{1/2}
\sum_{i=1}^6 [\int_{\Sigma_t^{out}}dV{\bf X}_i
{\bf X}_i]^{1/2}.
\label{3.16}
\end{equation}

Taking into account  estimates of the preceding
 subsection,
one obtains the inequality

\begin{equation}
 \partial_t H \le C_2(t)H+C_3(t),
\label{3.17}
\end{equation}

where $C_2, C_3$ are exponentially
bounded functions of coordinate time $t$
 with coeficients depending only on initial data.
(\ref{3.17}) readily implies

\begin{equation}
    H(t) \le
e^{\int_0^tds C_2(s) }\int_0^tdsC_3(s)
e^{-\int_0^tds C_2(s) }+
 H(t=0)e^{\int_0^tds C_2(s) }.
\label{3.18}
\end{equation}

thus proving the existence of a solution for any finite
time $t$. That ends the proof of Theorem 2.

\vskip 0.5cm
{\bf Remarks.}
The     Cauchy problem
should be investigated  in {\it weighted
Sobolev spaces}\cite{18} instead of Sobolev spaces.
The hypersurfaces $\Sigma_t^{out}$ are noncompact and
one should incorporate suitable falloff conditions
at spatial infinity; weighted Sobolev spaces include
them automatically, in contrast with the standard
Sobolev spaces. It is easy, however, to adapt the above
proof to work with $H_{s,\delta }$ instead of $H_s$ and
I will not discuss this point.
\vskip 2cm
\centerline{4. Final comments.}
 \par

 \par   The present investigation
 bases on the (1+3) splitting of spacetime that is
smooth (initially and possibly also globally, modulo a
region of compact support). The world time    $t$  can
  be  used globally to parametrize  casually related
occurences.   The proof that in electrovacuum event
horizons coincide with  apparent horizons is done with
only minimal reference to  specific properties of
spherically symmetric geometries. Obviously, the
existence of the maximal slicing requires a proof
\cite{14}, but the presence or absence of spherical
symmetry is probably of no great significance for the
validity of  maximal slicings. The   place where the
assumption of spherical symmetry   plays  important role
is the proof of the   identity (\ref{7})
but  it is quite likely that (\ref{7})
 survives (in the form of the Penrose-Gibbons
inequality) also  in nonspherical geometries.
Alternatively, one can use the Raychaudhuri
equation (\ref{15a}) in order to prove the local confining
property of apparent horizons, which   should  be of help
 in proving the existence of the global Cauchy solution.

The global Cauchy problem, however, poses a serious
 obstacle in
making a significant progress in proving the Cosmic
Censorship Hypothesis.

 The application of
the above ideas to  a more  general  class of spherically
symmetric geometries   of collapsing systems will
 be reported
elsewhere \cite{10}.

 \par Acknowledgements.
I wish to thank Niall O'Murchadha for
many discussions and valuable remarks, that greatly
contributed to my understanding of problems discussed
here. This work was supported in part by the Polish
Government Grant KBN 2526/93.
\vfill \eject
\centerline{\bf Appendix 1}

Static Einstein equations reduce to three equations
(\ref{a1}), (\ref{a3}) and (\ref{a5}), of which only
two are independent. Equation (\ref{a1}) gives the spatial
metric, which in a gauge $a(r)=b(r)=\phi^4$
 coincides with the
solution (\ref{1b}).  Inserting this  solution into
(\ref{a5}) gives (notice that $K_r^r=0$ and
$c={m^2-q^2\over 4}$)

$$\partial_r\ln \alpha = {\sqrt{a}\over p br^2}-{p\sqrt{a}
\over 4}-{q^2\sqrt{a}\over b^2r^4p}=$$

(here $p= {2(2\partial_r\phi r+\phi )\over \phi^3 r}$)

$$= {mr^2+4cr +mc\over  (r^2-c)(r^2+mr+c)}.$$

The last equation is solved by (\ref{1bb}).
\vskip 2cm
\centerline{\bf Appendix 2}

{\bf Lemma.} Under conditions stated in Theorem 2,
if  $\theta ' = p+K_r^r$ is positive on $\Sigma_0$
then it must be positive in the all
future  slices $\Sigma_t^{out}$.

{\bf Proof.} Assume the contrary, i. e., that there
exists a Cauchy (external) hypersurface $\Sigma_t^{out}$
such that somewhere on it $\theta '$ crosses through
zero.

The evolution of $\theta '$ is given by the equation

$$(\partial_t-{\alpha \over \sqrt{a}}\partial_r)\theta '
=$$

$$(-\partial_r\alpha /\sqrt{a} +\alpha K_r^r)\theta '
+\alpha \theta '^2/2.$$

 From this equation one infers that the
 surface with vanishing convergence
 $\theta '$  moves inward with the velocity of light
when immersed  in vacuum (Section 2). Therefore it must
 exist  in
all preceding  Cauchy slices and in particular in the
initial hypersurface $\Sigma_0^{out}$. This gives a
contradiction which proves our claim.

{\bf Corollary.} Assume that $\Sigma_0$ contains an
apparent horizon. Under conditions of the preceding
lemma, $C(t)$ (and, consequently, $K_r^r$)
must be strictly positive for any finite time $t$.

{\bf Proof.} From the results of Section 2, the
apparent horizon propagates to the future.
If $C(t)$ was equal to zero  on a slice
$\Sigma_t^{out}$  for some time $t$,
then both $\theta $ and $\theta '$ would
vanish at an apparent horizon, which would imply
(due to the
above lemma) the
existence of a white hole in $\Sigma_0$, contrary to the
assumption that $\theta ' >0$ in $\Sigma_0$.


\begin{references}
\bibitem{2} I am convinced that this form of the
electrovacuum (Reissner-Nordstroem) solution of Einstein
-Maxwell equations has been obtained hitherto -- its
space part is well known, see \cite{1} --  but I was not
able to find any reference in which the lapse function is
given.
\end{references}
 \end{document}